\documentclass[prl,showpacs,twocolumn]{revtex4}
\usepackage{graphics}
\usepackage{epsfig}
\usepackage{graphicx}

\begin{document}
\title{Entanglement swapping of the valence-bond solid state with local filtering operations}
\author{Heng Fan$^{1}$, Z. D. Wang$^2$, and Vlatko Vedral$^{3,4,5}$}
\affiliation{$^1$Institute of Physics, Chinese Academy of Sciences,
Beijing 100080, China \\
$^2$Department of Physics and
Center of Theoretical and Computational Physics,
The University of Hong Kong, Pokfulam Road, Hong Kong \\
$^3$The School of Physics and Astronomy, University of Leeds, Leeds,
LS2 9JT, United Kingdom \\
$^4$Centre for Quantum Technologies, National University of
Singapore, 3
Science Drive 2, Singapore 117543 \\
$^5$Department of Physics, National University of Singapore, 2
Science Drive 3, Singapore 117542 }

\date{\today}

\begin{abstract}
A basic problem of quantum communication is to generate entangled
states between distant sites. Using entanglement swapping, we are
able to generate an entangled state of the desired distance from
connecting many short distance entangled states. We investigate the
entanglement swapping of the Valence-Bond Solid state with
un-identical local filtering operations. It is found that not only a
long distance entangled state can be generated from the VBS state,
but also there is a trade-off between the probability and the degree
of the entanglement in the resultant state. The results are also
generalized to higher dimensional cases.
\end{abstract}

\maketitle

The goal of quantum communication is to transmit quantum states
between distant nodes. If a nearly perfect entangled state is shared
between distant nodes, a quantum state can be faithfully transferred
via quantum teleportation \cite{BBCJPW}. By entanglement swapping
\cite{ZZHE}, a maximally (perfect) entangled state of desired
distance can be generated from connecting short distance maximally
entangled states. For a chain of non-maximally entangled states, one
entangled state can be generated between two ends at the chain while
the degree of the entanglement decays exponentially with the number
of nodes when the connecting states are identical \cite{ACL}. The
exponential decay of the entanglement is a basic constraint for long
distance entanglement generation.  One may, however, overcome this
difficulty by various schemes depending on real physical systems
\cite{Briegel,DLCZ}, while essentially the entanglement with
exponential decay as a whole is divided into segments of {\it
off-line} entanglement preparation. Still the basic constraint of
entanglement with exponential decay in each segment exists.
Currently, almost all realistic quantum communication schemes are
based on photonic systems which are very attractive for long
distance quantum communication, such as over several hundred
kilometers. However, in a quantum computer based on solid states, it
is still necessary to transmit quantum states in nano-micron scale.
The Valence-Bond Solid (VBS) state is a fundamental solid state
system, which is well studied in the context of condensed matter
physics \cite{AKLT} and is found to be the ground state of the
gapped system \cite{Haldane}. More interestingly, the related states
may also be useful for universal quantum computation
\cite{Verstraete} and quantum computation by measurements
\cite{Miyake}. Yet the entanglement swapping implemented by the VBS
state has not been addressed. In this Letter, we show that a long
distance entangled state can be generated from the VBS state with
{\it local filtering operations} \cite{Gisin} by entanglement
swapping. A remarkable observation is that in entanglement swapping
of the VBS state the product of the degree of entanglement and the
probability to obtain this state is a constant, i.e., there is a
trade-off between the degree of entanglement and the probability of
the resultant state. This provides a basic constraint for
entanglement generation based on the VBS state. The result  can also
be generalized to higher dimension cases.

On the other hand, entanglement is not only a resource for quantum
information processing, but also it could lead to further insight
into other areas of physics. This has stimulated an exciting
cross-fertilization among those research areas, see Ref. \cite{AFAV}
for a review.
%The VBS state is the ground state of a gapped model
%\cite{AKLT,Haldane}.
The study of the VBS-type ground state  of
many-body systems with methods developed in quantum information may
unveil new properties. In particular, the study of entanglement
swapping for the VBS state is directly related to the localizable
entanglement and the string order parameter that reveals a hidden
topological long range order.

%In this Letter, we will study the entanglement swapping of the VBS
%state with local filtering operations. First, let's consider briefly
%the protocol of entanglement swapping.

%Letter,  This state is
%related with the `crackion' \cite{FS} and is a deformation of VBS
%state \cite{VMC,VR}.

%The study of entanglement of this state in this Letter is two-fold.
%On the one hand, we study entanglement-swapping (percolation) in
%quantum information processing with our unified approach. We will
%see that for a chain of entanglement swapping with different
%entangled pairs, the entanglement of the resulting state (measured
%by the concurrence) is the multiplication of all nearest-neighbor
%concurrences in the chain. This extends a recent result of
%\cite{ACL}. We also observe a trade-off between the entanglement of
%the resultant state and the corresponding probability. This
%constitutes a basic constraint for entanglement percolation. On the
%other hand, the localizable entanglement of the twisted VBS state
%can similarly be studied with our method. A rather rich structure of
%the string order operator is expected. In particular, we study the
%effect of a twisted operator on the boundaries, and contrary to
%common expectations, the boundary effects of a chain of identical
%twisted operators does not decay exponentially. We also consider the
%mutual information of the boundaries.

{\it Entanglement swapping protocol and the VBS state with local
filtering operations}-- We denote the maximally-entangled state as
$|\Phi ^+\rangle =(|00\rangle +|11\rangle )/\sqrt {2} $. The
entanglement-swapping relation \cite{ZZHE} takes the following form
which is similar to the well-known teleportation protocol
\cite{BBCJPW},
\begin{eqnarray}
|\Phi ^+\rangle _{AB}|\Phi ^+\rangle _{CD}= \frac {1}{2}\sum
_{i=0}^3 (I\otimes \sigma _i)|\Phi ^+\rangle _{AD}|\phi '_i\rangle
_{BC}, \label{swap}
\end{eqnarray}
where $\sigma _i$ are the Pauli matrices $\sigma _1=\sigma _x,\sigma
_2=\sigma _z, \sigma _3=\sigma _x\sigma _z$ and the identity matrix
$\sigma _0=I$ in two-dimension. Here we denote four Bell states as
$(I\otimes \sigma _i)|\Phi ^+\rangle \equiv |\phi '_i\rangle$, which
are four orthogonal maximally entangled states.

We now recall  the local filtering operator  $T_j=\sqrt{2}(\alpha
_j|0\rangle \langle 0|+\beta _j|1\rangle \langle 1|)$ (with $|\alpha
_j|^2+|\beta _j|^2=1$) introduced in Ref.\cite{Gisin}, which in
general is not unitary and 'squashes' the entanglement.  The state
$|\Phi _j\rangle \equiv \alpha _j|00\rangle +\beta _j|11\rangle $
can then be written in terms of the local filtering operator as
$|\Phi _j\rangle =(I\otimes T_j)|\Phi ^+\rangle $. For two entangled
pairs with the corresponding local filtering operators, the
entanglement-swapping relation gives
\begin{eqnarray}
|\Phi _0\rangle _{AB}|\Phi _1\rangle _{CD}= \frac {1}{2}\sum
_{i=0}^3 (I\otimes T_1\sigma _iT_0)|\Phi ^+\rangle _{AD}|\phi '
_i\rangle _{BC}, \label{swap1}
\end{eqnarray}
where we have additionally used the property $(T_j\otimes I)|\Phi
^+\rangle =(I\otimes T_j)|\Phi ^+\rangle $. Performing the
Bell-state measurement at nodes $BC$, the resulting states between
nodes $A$ and $D$ are $(I\otimes T_1\sigma _iT_0)|\Phi ^+\rangle
_{AD}$ with a normalization factor. Here we would like to point out
that if a local filtering operator $T'$ has off-diagonal entries, it
can be changed to a diagonal operator $T$ as $T=UT'V$ where $U$ and
$V$ are unitary. From entanglement theory, we know that this
transformation does not change the entanglement.

Now let us consider a generalized VBS state,
\begin{eqnarray}
|V\rangle =\prod _{j=0}^N( \alpha _{j}a_j^{\dagger }b_{j+1}^{\dagger
}-\beta _{j}b_j^{\dagger }a_{j+1}^{\dagger })|vac\rangle ,
\label{fvbs}
\end{eqnarray}
where $j$ is the lattice site, $a_j^{\dagger }$ and $b_j^{\dagger }$
are bosonic operators satisfying $[a^{\dagger }_i,b^{\dagger }_j]=
[a^{\dagger }_i,a^{\dagger }_j]=[b^{\dagger }_i,b^{\dagger }_j]=0$.
$|vac\rangle $ is the vacuum state. For $\alpha _j=\beta _j
=1/\sqrt{2}$, the state $|V\rangle $ is the standard spin-1 VBS
state which is the ground state of the gapped model (AKLT)
\cite{AKLT}. The state (\ref{fvbs}) can be considered as a
generalization of the VBS state, it changes continuously from the
ferromagnetic state to VBS state by changing parameters $\alpha_j$
and $\beta_j$. The state (\ref{fvbs}) is also the ground state of
the AKLT model with chiral parameters.

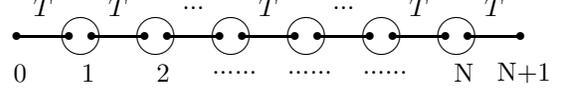
\begin{center}
\unitlength=1mm
\begin{picture}(80,5)(0,3)
\put(2,0){\makebox(3,3){$T$}} \put(12,0){\makebox(3,3){$T$}}
\put(22,0){\makebox(3,3){$...$}} \put(32,0){\makebox(3,3){$T$}}
\put(42,0){\makebox(3,3){$...$}} \put(52,0){\makebox(3,3){$T$}}
\put(62,0){\makebox(3,3){$T$}}
\end{picture}
\begin{picture}(80,5)(0,0)
\put(0,0){\line(1,0){7}} \put(10,0){\line(1,0){7}}
\put(20,0){\line(1,0){7}} \put(30,0){\line(1,0){7}}
\put(40,0){\line(1,0){7}} \put(50,0){\line(1,0){7}}
\put(60,0){\line(1,0){7}} \put(0,0){\circle*{1}}
\put(7,0){\circle*{1}} \put(10,0){\circle*{1}}
\put(17,0){\circle*{1}} \put(20,0){\circle*{1}}
\put(27,0){\circle*{1}} \put(30,0){\circle*{1}}
\put(37,0){\circle*{1}} \put(40,0){\circle*{1}}
\put(47,0){\circle*{1}} \put(50,0){\circle*{1}}
\put(57,0){\circle*{1}} \put(60,0){\circle*{1}}
\put(67,0){\circle*{1}} \put(8.5,0){\circle{5}}
\put(18.5,0){\circle{5}} \put(28.5,0){\circle{5}}
\put(38.5,0){\circle{5}} \put(48.5,0){\circle{5}}
\put(58.5,0){\circle{5}}
\end{picture}
\begin{picture}(80,5)(1,1)
\put(0,0){\makebox(3,3){0}} \put(9,0){\makebox(3,3){1}}
%\put(10,0){\makebox(3,3){$\bar {1}$}}
\put(19,0){\makebox(3,3){2}}
%\put(20,0){\makebox(3,3){$\bar {2}$}}
\put(27,0){\makebox(3,3){...}} \put(30,0){\makebox(3,3){...}}
\put(37,0){\makebox(3,3){...}} \put(40,0){\makebox(3,3){...}}
\put(47,0){\makebox(3,3){...}} \put(50,0){\makebox(3,3){...}}
\put(59,0){\makebox(3,3){N}}
%\put(60,0){\makebox(3,3){$\bar {N}$}}
\put(67,0){\makebox(3,3){N+1}}
\end{picture}
\end{center}
%\medskip
A schematic diagram for a chain of entangled states: two black dots
connected by a line represent an entangled state with a local
filtering operator $T$, and circles denote the projection onto the
symmetric subspace of spin-1. By Bell measurement on each circle, an
entangled state is shared between nodes $0, N+1$.\\

We start from a simple case with only two terms in (\ref{fvbs}), for
instance $j=0,1$,
\begin{widetext}
\begin{eqnarray}
&&( \alpha _0a_0^{\dagger }b_{1}^{\dagger }-\beta _0b_0^{\dagger
}a_{1}^{\dagger })( \alpha _1a_1^{\dagger }b_{2}^{\dagger }-\beta
_1b_1^{\dagger }a_{2}^{\dagger }) |vac\rangle \nonumber \\
%&=&(\alpha _0\alpha _1a^{\dagger }_0b^{\dagger }_2+\beta _0\beta
%_1b^{\dagger }_0a^{\dagger }_2)a^{\dagger }_1b^{\dagger
%}_1|vac\rangle -\sqrt {2}\alpha _0\beta _1a^{\dagger }_0a^{\dagger
%}_2\frac {(b^{\dagger }_1)^2}{\sqrt {2}}|vac\rangle-\sqrt {2}\alpha
%_1\beta _0b^{\dagger }_0b^{\dagger }_2\frac {(a^{\dagger
%}_1)^2}{\sqrt {2}}|vac\rangle
%\nonumber \\
&=& (\alpha _0\alpha _1a^{\dagger }_0b^{\dagger }_2+\beta _0\beta
_1b^{\dagger }_0a^{\dagger }_2)a^{\dagger }_1b^{\dagger
}_1|vac\rangle -\frac {1}{\sqrt {2}}(\alpha _0\beta _1a^{\dagger
}_0a^{\dagger }_2+\alpha _1\beta _0b^{\dagger }_0b^{\dagger
}_2)(\frac {(a^{\dagger }_1)^2}{\sqrt {2}}+ \frac {(b^{\dagger
}_1)^2}{\sqrt
{2}})|vac\rangle  \nonumber \\
&&+ \frac {1}{\sqrt {2}}(\alpha _0\beta _1a^{\dagger }_0a^{\dagger
}_2-\alpha _1\beta _0b^{\dagger }_0b^{\dagger }_2) (\frac
{(a^{\dagger }_1)^2}{\sqrt {2}}- \frac {(b^{\dagger }_1)^2}{\sqrt
{2}})|vac\rangle . \label{fvbscal}
\end{eqnarray}
\end{widetext}
We notice that the lattice site $1$ is in a three-dimensional Fock
space in the bosonic representation as $a^{\dagger }_1b^{\dagger
}_1|vac\rangle ,\frac {(a_1^{\dagger })^2}{\sqrt {2}}|vac\rangle $
and $\frac {(a_1^{\dagger })^2}{\sqrt {2}}|vac\rangle $. We may,
however, represent the Fock space by two-qubit states as $a^{\dagger
}_1b^{\dagger }_1|vac\rangle \equiv (|01\rangle +|10\rangle )$
,$\frac {(a_1^{\dagger })^2}{\sqrt {2}}|vac\rangle \equiv |00\rangle
$, $\frac {(a_1^{\dagger })^2}{\sqrt {2}}|vac\rangle \equiv
|11\rangle $, here we have used the notations $a^{\dagger
}_j|vac\rangle =|0\rangle _j$ and $b^{\dagger }_j|vac\rangle
=|1\rangle _j$. Note that there is no singlet state (antisymmetric
state). Then $( \alpha _ja_j^{\dagger }b_{j+1}^{\dagger }-\beta
_jb_j^{\dagger }a_{j+1}^{\dagger })|vac\rangle =\alpha _j|01\rangle
-\beta _j|10\rangle $, which can also be written as $(I\otimes
\sigma _3T_j)|\Phi ^+\rangle =|\Psi _j\rangle $. Since the local
filtering operators are present in the equation, we thus call state
(\ref{fvbs}) as the VBS state with local filtering operations.

It is observed that the result in (\ref{fvbscal}) is like the
entanglement swapping with excluding the antisymmetric state.
Nevertheless, similar to the entanglement swapping (\ref{swap1}), we
can still reformulate Eq.(\ref{fvbscal}) as,
\begin{eqnarray}
\prod _{j=0}^1 &&( \alpha _ja_j^{\dagger }b_{j+1}^{\dagger }-\beta
_jb_j^{\dagger }a_{j+1}^{\dagger })|vac\rangle \nonumber
\\
&=&S_{BC}|\Psi
_0\rangle _{AB}|\Psi _1\rangle _{CD} \nonumber \\
&=& \sum _{i=1}^3 (I\otimes \sigma _3T_1\sigma _iT_0)|\Phi ^+\rangle
_{AD}|\phi _i\rangle _{BC}, \label{swap2}
\end{eqnarray}
where we have used the notation $|\phi _i\rangle =(\sigma _3\otimes
\sigma _i)|\Phi ^+\rangle $, $A$ and $D$ are respectively the
lattice sites $0$ and $2$, and both $B$ and $C$ correspond to the
lattice site $1$ in (\ref{fvbscal}). The essential difference
between Eq. (\ref{swap2}) and Eq. (\ref{swap1}) is the operator
$S_{BC}$ that projects the two-qubit space to its three-dimensional
symmetric subspace, i.e., the singlet state is projected out.
Consequently, the summation in (\ref{swap2}) goes from 1 to 3,
rather than 0 to 3. Note that the normalization factor is omitted in
Eq. (\ref{swap2}).

By repeatedly using this result, the VBS state with local filtering
operations (\ref{fvbs}) can be formulated as
\begin{eqnarray}
|V\rangle =\frac {1}{\sqrt{P_{tot}}}\sum _{i_1,...,i_N=1}^3(I\otimes
\overrightarrow{\cal {T}}_{i_1...i_N})|\Phi ^+\rangle _{0N+1}|\phi
_{i_1}...\phi _{i_N}\rangle , \label{swapchain1}
\end{eqnarray}
where $\overrightarrow{\cal {T}}_{i_1...i_N}\equiv \sigma
_3T_{N}\sigma _{i_N}T_{N-1}...\sigma _{i_1}T_0$, and
$1/\sqrt{P_{tot}}$ is the normalization factor to be explicitly
given below.

{\it Entanglement swapping of the VBS state: General results}--Let
us begin with the simplest case as in (\ref{swap2}). By Bell
measurement operation in node $(B,C)$ with $\{|\phi \rangle _i\}
_{i=1}^3$, the states shared between nodes $A$ and $D$ are: $(\alpha
_0\alpha _1a^{\dagger }_0b^{\dagger }_2+\beta _0\beta _1b^{\dagger
}_0a^{\dagger }_2)|vac\rangle $, $(\alpha _0\beta _1a^{\dagger
}_0a^{\dagger }_2+\alpha _1\beta _0b^{\dagger }_0b^{\dagger
}_2)|vac\rangle $ and $(\alpha _0\beta _1a^{\dagger }_0a^{\dagger
}_2-\alpha _1\beta _0b^{\dagger }_0b^{\dagger }_2)|vac\rangle $ with
respectively probability $|\alpha _0\alpha _1|^2+|\beta _0\beta
_1|^2$, $|\alpha _0\beta _1|^2+|\beta _0\alpha _1|^2$ and $|\alpha
_0\beta _1|^2+|\beta _0\alpha _1|^2$,  while the resultant
entanglement quantified by concurrence, a well-accepted entanglement
measure \cite{W}, takes the form $|\alpha _0\beta _0\alpha _1\beta
_1|/(|\alpha _0\alpha _1|^2+|\beta _0\beta _1|^2)$, $|\alpha _0\beta
_0\alpha _1\beta _1|/ (|\alpha _0\beta _1|^2+|\beta _0\alpha _1|^2)$
and $|\alpha _0\beta _0\alpha _1\beta _1|/(|\alpha _0\beta
_1|^2+|\beta _0\alpha _1|^2)$ correspondingly. Our observation is
that for each resultant state, the product of the probability and
the corresponding concurrence equals to a constant $|\alpha _0\beta
_0|\times |\alpha _1\beta _1|$, which is just  the product of the
two concurrences corresponding to the original valence-bond solid
states $( \alpha _0a_0^{\dagger }b_{1}^{\dagger }-\beta
_0b_0^{\dagger }a_{1}^{\dagger })|vac\rangle $ and $( \alpha
_1a_1^{\dagger }b_{2}^{\dagger }-\beta _1b_1^{\dagger
}a_{2}^{\dagger })|vac\rangle $. For an ideal case where the
original states shared are maximally entangled states, we can always
obtain a maximally entangled state shared between nodes $A$ and $D$
with the probability one. While for a general case, if the
entanglement is large, the probability of obtaining it is small, and
vise versa. This is an interesting result: for entanglement swapping
of the VBS state with local filtering operations, there is a
trade-off for the resultant state between the probability and the
degree of the entanglement; their product is uppermost bounded by
the degree of original entanglement. The reason to have such upper
bound lies in that other measurements different from the Bell
measurement generally diminish the final entanglement.

Considering the VBS state in (\ref{swapchain1}), by a sequential of
Bell measurement in nodes $1,...,N$ corresponding to the measurement
outcome $|\phi _{i_1}\rangle ,..., |\phi _{i_N}\rangle $ with
probability ${\rm \widetilde{Prob}}(i_1...i_N)$, the final resultant
state shared between the two end nodes is
\begin{eqnarray}
|\psi \rangle _{{0N+1}}=(I\otimes \overrightarrow{\cal
{T}}_{i_1...i_N})|\Phi ^+\rangle _{0N+1}/\sqrt{P_{i_1...i_N}}.
\end{eqnarray}
Here the probability to obtain this state is ${\rm
\widetilde{Prob}}(i_1...i_N)=P_{i_1...i_N}/P_{sum}$, with
$P_{i_1...i_N}=\frac {1}{2}{\rm Tr}( T_{N}...\sigma
_{i_1}T_0T_0^{\dagger } \sigma ^{\dagger }_{i_1}...T^{\dagger
}_{N})$ and
%$P_{sum}$ is a summation
$P_{sum}=\sum _{i_1...i_N=1}^3P_{i_1...i_N}$ as a normalization
factor. (Of course, the summation of all probabilities equals to
one: $\sum P_{i_1...i_N}=1$). Note that the measurement result
corresponding to the singlet is ruled out. The concurrence of the
final state $|\psi _{0N+1}\rangle $ defined as ${\cal
{C}}_{i_1...i_N}=2\sqrt{{\rm det}(\rho _{N+1})}$, with $\rho _{N+1}$
as the reduced density operator, is found as
\begin{eqnarray}
{\cal {C}}_{i_1...i_N}=\prod _{j=0}^{N}{\cal {C}} _j/P_{i_1...i_N},
\end{eqnarray}
where ${\cal {C}}_j=2|\alpha _j\beta _j|$ is the concurrence of the
$j$-th valence-bond state $( \alpha _{j}a_j^{\dagger
}b_{j+1}^{\dagger }-\beta _{j}b_j^{\dagger }a_{j+1}^{\dagger
})|vac\rangle $. At this stage, we establish an explicit relation
 between the probability of obtaining a
resultant state and its degree of entanglement:
\begin{eqnarray}
{\rm \widetilde{Prob}}(i_1...i_N){\cal {C}}_{i_1...i_N}=\prod
_{j=0}^{N}{\cal {C}} _j/P_{sum},
\end{eqnarray}
noting that the RHS of the above equation is a quantity independent
of the measurement details. If the probability is large, the
corresponding entanglement is small, and vise versa. As a result,
although a long distance entangled state can be generated from the
VBS state with local filtering operations,  one can only increase
either the probability or the degree of the entanglement for the
resultant state, but not both if the original entanglement of each
valence bond ${\cal {C}}_j$ is fixed.

{\it Entanglement swapping of the VBS state: Examples}-- We consider
a special case where all local filtering operators are the same
$T_j=T=\sqrt{2}(\alpha |0\rangle \langle 0|+\beta |1\rangle \langle
1|)$. In this case, the entanglement of each valence-bond state $(
\alpha a_j^{\dagger }b_{j+1}^{\dagger }-\beta b_j^{\dagger
}a_{j+1}^{\dagger })|vac\rangle $ equals to ${\cal {C}}=2|\alpha
\beta |$. By entanglement swapping, we have
\begin{eqnarray}
{\rm \widetilde{Prob}}(i_1...i_N){\cal {C}}_{i_1...i_N}={\cal
{C}}^{N+1} /P_{sum}.
\end{eqnarray}
If the original entanglement of the valence-bond states are not
maximally entangled, %the final resultant entanglement will
either the probability or the degree of the entanglement decays
exponentially. For example, if $N$ is even, by some calculations, it
is possible that the final state shared between two ends is a
maximally entangled state: $(a_0^{\dagger }b_{N+1}^{\dagger
}-b_0^{\dagger }a_{N+1}^{\dagger })|vac\rangle $, meanwhile the
probability to obtain this state decays exponentially with the
number of bonds $N$, i.e., ${\rm \widetilde{Prob}}(i_1...i_N)={\cal
{C}}^{N+1} /P_{sum}$ simply because the entanglement of the final
state  ${\cal {C}}_{i_1...i_N}=1$. For the case where no local
filtering operators are present, we know that the final entanglement
is always one, i.e.,  the final state is one of the three Bell
states (exclusive the singlet state) exactly with the probability
$1/3$. The entanglement swapping of identical non-maximally
entangled states in a chain and a network was also addressed before
\cite{ACL}.

{\it Entanglement swapping in higher dimension with local filtering
operations}-- The generalized Pauli matrices in $D$-dimension are
defined as $Z=\sum _j\omega ^j|j\rangle \langle j|, X=\sum
_j|j+1{\rm mod}D\rangle \langle j|$, where $\omega =e^{2\pi i/D}$,
$\{|j\rangle \}_{j=0}^{D-1}$ is an orthonormal basis, and
$\{U_{mn}=X^mZ^n\} _{m,n=0}^{D-1}$ constitutes a basis of unitary
operators in d-dimensional Hilbert space. Defining $|\Phi ^+\rangle
=\frac {1}{\sqrt{D}}\sum _{j=0}^{D-1}|jj\rangle $, it is known that
$\{|\Phi _{mn}\rangle =(I\otimes U_{mn})|\Phi ^+\rangle
\}_{m,n=0}^{D-1}$ constitutes the orthonormal bases of the maximally
entangled states, where $I$ is the identity operator in
$D$-dimensional Hilbert space.  We introduce the local filtering
operators as $T_k=\frac {1}{D}\sum _{j=0}^{D-1}\lambda ^k_j|j\rangle
\langle j|$, and  still denote $|\Phi _k\rangle =(I\otimes T_k)|\Phi
^+\rangle $ correspondingly. By repeatedly using entanglement
swapping protocol as in, for example, Ref.\cite{Fan}, $|\Phi
^+\rangle _{AB}|\Phi ^+\rangle _{CD} =\frac {1}{D}\sum _{m,n}|\Phi
_{m,n}\rangle _{AC}|\Phi _{m,-n}\rangle _{BD}$, we have
\begin{eqnarray}
\prod _{k=0}^N|\Phi _k\rangle =\frac {1}{D^N}\sum
_{m_k,n_k}(I\otimes \widetilde{T})|\Phi ^+\rangle _{0,N+1}\prod
_k|\Phi _{m_k,-n_k}\rangle ,
\end{eqnarray}
where $\widetilde{T}=T_NU_{m_Nn_N}T_{N-1}...U_{m_1n_1}T_0$, and
 the relation $(I\otimes
T)|\Phi ^+\rangle =(T\otimes I)|\Phi ^+\rangle $ has been used,
similar to that in the two-dimensional case. By using a generalized
Bell measurement, the final state shared between two end nodes is a
pure state $(I\otimes \widetilde{T})|\Phi ^+\rangle _{0,N+1}$.

If we adopt a $D$-dimensional concurrence definition given by ${\cal
{C}}=\sqrt{1-{\rm Tr}\rho _{N+1}}$ (up to a factor),  we can hardly
obtain an explicit simple expression for the final entanglement.
Fortunately, a class of entanglement measures are available, and we
here choose a generalized concurrence in a hierarchy
\cite{Fan1,Gour} to evaluate the entanglement, which is
%In this Letter, we consider an
%extended concurrence
defined as ${\cal {C}}^e=D\sqrt[n]{{\rm det}\rho _{N+1}}$, where
$\rho _{N+1}$ is the reduced density operator of the state
$(I\otimes \widetilde{T})|\Phi ^+\rangle _{0,N+1}$, noting that
$\widetilde{T}$ depends on $m_k,n_k$. Since the entanglement of the
state $|\Phi _k\rangle =(I\otimes T_k)|\Phi ^+\rangle $ is ${\cal
{C}}^e_k=D\sqrt[n]{{\rm det}T_k}$, the resultant entanglement
between two ends is,
\begin{eqnarray}
{\cal {C}}^e=\prod _{k=0}^N{\cal
{C}}^e_k/\widetilde{Prob}(m_j,n_j,j=0...N), \label{gswap}
\end{eqnarray}
where $\widetilde{Prob}(m_j,n_j,j=0...N)$ is the probability
corresponding to $\widetilde{T}$. Thanks to the relation ${\rm
det}U_{mn}=1$ such that we have obtained this simple and concise
equation. If all local filtering operations are the same, the
product of the final entanglement and the corresponding probability
decays exponentially with the number of nodes $N$. For an extreme
case, if all states $|\Phi _k\rangle $ are maximally entangled, we
 can always generate a maximally entangled state shared between
distance two ends.

For a generalized VBS state, we can consider a projection operation
on each nodes ${\cal {P}}\prod _{k=0}^N|\Phi _k\rangle $. Similar to
that for the two-dimensional case, the entanglement swapping can be
constructed, and a whole factor will appear in (\ref{gswap}).

{\it Discussion and Summary}--While the VBS state and the related
spin-1 chain were extensively studied before, here we have explored
for the first time  the entanglement swapping based on this system
with non-identical local filtering operations. The local filtering
operators may be adjusted according to the real physical systems.
The Bell measurement for solid state systems can be realized by
electron-pair beams splitter setup \cite{Loss}, the similar setup is
standard in quantum optics. If each valence bound is identical and
not maximally entangled, then the resultant entanglement decays
exponentially with the length of the chain. Since a long distance
entanglement is a precious resource for quantum information
processing, we may try to overcome the exponential decay by dividing
the chain into several segments with each segment length comparable
with the whole length. The entanglement swapping can be performed
off-line in each segment, and if succeed, we can connect them
together to constitute a longer one. We also wish to remark that the
approach developed here is a unified one, which can also be applied,
for example, in  entanglement purification experiments where not all
primitive entangled pair photons are equal or maximally entangled.

Similar to the qubit case,  the high dimension entanglement swapping
should also have the  exponential decay of entanglement for a chain
of non-maximally entangled states. We have shown this  result
explicitly by using a generalized concurrence for quantifying
entanglement.
 %to quantify the entanglement.

{\it Acknowledgements}, HF acknowledges the support by "Bairen"
program, NSFC grant (10674162) and "973" program (2006CB921107). ZDW
thanks to the support of Hong Kong GRF (HKU7051/06P and
HKU7044/08P). VV acknowledges the Royal Society, the Wolfson Trust,
the Engineering and Physical Sciences Research Council (UK) and the
National Research Foundation and Ministry of Education (Singapore)
for their financial support.


\begin{thebibliography}{99}
\bibitem{BBCJPW}C.H.Bennett, G.Brassard, C.Crepeau,
R.Jozsa, A.Peres, W.K.Wootters, Phys.Rev.Lett. {\bf 70}, 1895(1993).

\bibitem{ZZHE}M. Zukowski,
A. Zeilinger, M. A. Horne, and A. Ekert, Phys. Rev. Lett. {\bf 71},
4287 (1993).

\bibitem{ACL}A. Acin, J. I. Cirac, M. Lewenstein, Nature Phys. {\bf 3}, 256 ( 2007).


\bibitem{Briegel}H. J. Briegel, W. Duer, J. I. Cirac, and P. Zoller,
Phys. Rev. Lett. {\bf 81}, 5932 (1991).

\bibitem{DLCZ}L. M. Duan,
M. D. Lukin, J. I. Cirac, and P. Zoller, Nature {\bf 414}, 413
(2001).

\bibitem{AKLT}A. Affleck, T. Kennedy, E. H. Lieb, and H. Tasaki,
Phys. Rev. Lett. {\bf 59}, 799 (1987); Commun. Math. Phys. {\bf
115}, 477 (1988).

\bibitem{Haldane}F. D. M. Haldane, Phys. Rev. Lett. {\bf 50},
1153 (1983).



\bibitem{Verstraete}F. Verstraete, J. I. Cirac, Phys. Rev. {\bf A}
70, 060302 (2004).

\bibitem{Miyake}G. K. Brennen, A. Miyake, Phys. Rev. Lett. {\bf
101}, 010502 (2008).



\bibitem{Gisin}N. Gisin, Phys. Lett. {\bf A} 210, 151 (1996).

\bibitem{AFAV}L. Amico, R. Fazio, A. Osterloh, and V. Vedral,
Rev. Mod. Phys. {\bf 80}, 517 (2008).

\bibitem{W}W. K. Wootters, Phys. Rev. Lett. {\bf 78}, 5022 (1997).

\bibitem{Fan}H. Fan, Phys. Rev. Lett. {\bf 92}, 177905 (2004).

\bibitem{Fan1}H. Fan, J. Phys. A {\bf 36}, 4151 (2003).

\bibitem{Gour}G. Gour, Phys. Rev. A {\bf 71}, 012318 (2005).

%\bibitem{FKR}H. Fan
%V. Korepin, and V. Roychowdhury, Phys. Rev. Lett. {\bf 93}, 227203
%(2004).



\bibitem{Loss}G. Burkard, and D. Loss, Phys. Rev. Lett. {\bf 91},
087903 (2003) .


%\bibitem{DKH}R. Dillenschneider {\it et al.}, arXiv:0705.3993;
%T.Hirano {\it et al}, arXiv:0710.4198.

%\bibitem{FS}G.Fath and J.S\'olyom, J.Phys:Cond. Matt.{\bf
%5},8983(1993); K. Totsuka and M. Suzuki, J.Phys:Cond. Matt., {\bf
%7}, 1639 (1995).



\end{thebibliography}
\end{document}